\documentclass{article}
\usepackage{itw2004}
\usepackage{amssymb, amsthm, amsmath}

\usepackage{epsfig,bbm}

\input psfig.sty

\newtheorem{theorem}{Theorem}[section]
\newtheorem{lemma}[theorem]{Lemma}

\newtheorem{corollary}[theorem]{Corollary}

\theoremstyle{definition}
\newtheorem{definition}[theorem]{Definition}
\newtheorem{example}[theorem]{Example}

\theoremstyle{definition}  
\newtheorem{remark}[theorem]{Remark}

\numberwithin{equation}{section}

\def\Z{{\mathbb{Z}}}
\def\N{{\mathbb{N}}}

\def\del{{\partial}}
\newcommand{\R}{\mathbb{R}}

\newcommand{\matr}[1]{#1}
\newcommand{\vect}[1]{\mathbf{#1}}
\newcommand{\set}[1]{#1}
\newcommand{\code}[1]{#1}
\newcommand{\graph}[1]{#1}

\newcommand{\tvc}{\vect{\widetilde c}}

\newcommand{\vx}{\vect{x}}

\newcommand{\dirX}{\vec X}
\newcommand{\dirN}{\vec N}
\newcommand{\dirV}{\vec V}
\newcommand{\dirE}{\vec E}

\newcommand{\vomega}{\boldsymbol{\omega}}

\newcommand{\fch}[2]{{#1}(\matr{#2})}

\newcommand{\dmin}{d_{\mathrm{min}}}

\newcommand{\matrunity}{I}

\newcommand{\GF}[1]{\mathbb{F}_{#1}}

\newcommand{\defeq}{\triangleq}

\begin{document}


\topmargin = 0mm


\itwtitle{Pseudo-Codewords of Cycle Codes via Zeta Functions%
  \footnote[0]{dummytext}}


\itwauthor{Ralf Koetter\footnote{dummytext}}
{Coordinated Science Lab. \\
University of Illinois \\
Urbana, IL  61801 \\
{\tt koetter@uiuc.edu}
}

\itwsecondauthor{Wen-Ching W.\ Li\footnote{dummytext}}
{Department of Mathematics\\
Pennsylvania State University\\
University Park, PA 16802-6401\\
\tt{wli@math.psu.edu}
}

\itwthirdauthor{Pascal O.\ Vontobel\footnote{dummytext}}
{Coordinated Science Lab. \\
University of Illinois \\
Urbana, IL  61801 \\
\tt{vontobel@ifp.uiuc.edu}
}

\itwfourthauthor{Judy L.\ Walker\footnote{dummytext}}
{Department of Mathematics\\
University of Nebraska\\
Lincoln, NE  68588-0130\\
{\tt jwalker@math.unl.edu}
}


\itwmaketitle

 
\footnotetext[0]{This is essentially the paper that was presented at the IT
                 Workshop 2004, San Antonio, TX, USA. We replaced ``Newton
                 polytope'' by ``Newton polyhedron'' throughout the text and
                 corrected a slight unpreciseness in
                 Th.~\ref{theorem:pseudos:in:cycle:1}}

\footnotetext[1]{R.~Koetter's research was partially supported by NSF Grants
                 CCR 99-84515 and CCR 01-05719.}

\footnotetext[2]{W.-C.~W.~Li's research was partially supported by NSA Grant
                 MDA904-03-1-0069.}

\footnotetext[3]{P.~O.~Vontobel is now with the ECE Dept., University of
                 Wisconsin-Madison, USA, {\tt vontobel@ece.wisc.edu}; his
                 research was partially supported by NSF Grants CCR 99-84515
                 and CCR 01-05719.}

\footnotetext[4]{J.~L.~Walker's research was partially supported by NSF Grant
                 DMS 03-02024.}


\begin{itwabstract} 
  Cycle codes are a special case of low-density parity-check (LDPC) codes and
  as such can be decoded using an iterative message-passing decoding algorithm
  on the associated Tanner graph. The existence of pseudo-codewords is known
  to cause the decoding algorithm to fail in certain instances. In this paper,
  we draw a connection between pseudo-codewords of cycle codes and the
  so-called edge zeta function of the associated normal graph and show how the
  Newton polyhedron of the zeta function equals the fundamental cone of the
  code, which plays a crucial role in characterizing the performance of
  iterative decoding algorithms.
\end{itwabstract}


\begin{itwpaper}


\vspace{-1mm}

\itwsection{Introduction}

\vspace{-1mm}

We are interested in characterizing the performance of a binary
low-density parity-check (LDPC) code $\code{C}$ used for the
transmission of information over a memoryless channel. Moreover, we
focus on the case that iterative decoding is performed at the receiver
end.

Let the code be described by a parity-check matrix $\matr{H}$. To a
matrix $H$ we can associate a bipartite graph, the so-called Tanner
graph $T \defeq T(\matr{H})$~\cite{Tanner:81}. 
As was realized in~\cite{Koetter:Vontobel:03:1}, an essential role in
the understanding of iterative decoding is played by the finite covers
of the Tanner graph $T$ and the codes defined by them. In fact, while
the main strength of iterative decoders, namely their low complexity,
results from the fact that they operate {\em locally} on the Tanner
graph, this very fact is also the source of the weakness of any
iterative decoding algorithm. The systemic problem is that by just
performing local operations the decoder cannot distinguish if it is
decoding on the Tanner graph $T$ or any of the finite covers. Thus,
codewords in a cover of $T$ will be interfering with the iterative
decoding process.  Consequently, in order to understand the behavior
of iterative decoders we will have to characterize the ``covering''
codes and their codewords.

The goal of this paper is to give a concise geometric and simple
description of these codes in finite covers of $T$.  In particular,
the geometric characterization will relate to a cone in Euclidean
space, the so-called fundamental cone~\cite{Koetter:Vontobel:03:1}.


We focus on a special class of LDPC codes, namely the class of cycle
codes. These codes are informally defined as LDPC codes where all bit
nodes  have degree two.\footnote{The reason for the
  name ``cycle codes'' will become clear in
  Sec.~\ref{sec:binary:linear:codes:1}.}

From a practical point of view cycle codes are somewhat marred by the
fact that their minimum distance grows (at best) logarithmically in
the block length (assuming fixed check-node degrees). Nevertheless,
their properties make them more amenable to analysis than general LDPC
codes.  In any case, cycle codes can be seen as an interesting object
of study from which results can (hopefully) be suitably generalized to
the more interesting class of LDPC codes where part or all of the bit
nodes have degree at least three.

The connections between iterative decoding and LDPC codes are probably best
understood for cycle codes. First of all, the fundamental cone can be related
concisely to the decoding behavior under iterative decoding, and secondly, as
we aim to show in this paper, the fundamental cone may be identified as the
Newton polyhedron of Hashimoto's edge zeta function \cite{Hashimoto:89:1}
associated to the normal graph (defined in
Sec.~\ref{sec:binary:linear:codes:1}) of the code.  For an early reference
about the performance of iterative decoding techniques in conjunction with
cycle codes see \cite[ch.6]{nicwi}.  In the case of general LDPC codes, the
relation of the fundamental cone to the exact characterization of the
iterative decoding behavior is more intricate. Nevertheless, even here the
fundamental cone gives an amazingly exact picture of the
behavior.\footnote{The behavior of the linear programming
decoder~\cite{Feldman:Karger:Wainwright:03:2} (for the most canonical
relaxation) is \emph{exactly} characterized by the fundamental cone in the
cycle code case \emph{and} in the non-cycle code case.}  While we here only
establish the connection between the fundamental cone and the edge zeta
function for cycle codes, we conjecture the existence of such a zeta function
for the case of general LDPC codes.

This paper is structured as follows: Sec.~\ref{sec:binary:linear:codes:1}
introduces the basics about Tanner graphs and normal graphs of binary linear
codes and Sec.~\ref{sec:fundamental:cone} discusses graph covers and the
fundamental cone associated with a code. The notion of an edge zeta function
of a graph will be introduced in Sec.~\ref{sec:zeta:functions:1} and
Sec.~\ref{sec:connection:fundamental:cone:zeta:function:1} discusses the main
result of this paper, namely the identification of the fundamental cone and
the Newton polyhedron in the case of cycle codes. Throughout the whole paper
we will use two running examples containing two different codes, namely Code A
and Code B: the first is \emph{not} a cycle code whereas the latter one is a
cycle code.


\itwsection{Binary Linear Codes and Their Graphs}

\label{sec:binary:linear:codes:1}


\begin{definition}
  \label{def:undirected:graph:1}

  An \emph{undirected graph} $\graph{X} =
  \graph{X}(\set{V}(\graph{X}), \set{E}(\graph{X}))$ consists of a
  vertex-set $\set{V} \defeq \set{V}(\graph{X})$ and an edge-set
  $\set{E} \defeq \set{E}(\graph{X})$ where the elements of $\set{E}$
  are 2-subsets of $\set{V}$. We assume a fixed ordering on $E$ so
  that $E = \{e_1, \dots, e_n\}$, where $n \defeq n(X) \defeq |E|$. By
  a {\em graph} (without further qualifications) we will always mean an
  undirected graph. We will not allow self-loops or multiple
  edges. For $v \in V$, we write $\del(v)$ for the {\em neighborhood}
  of $v$, i.e., the collection of vertices of $X$ which are adjacent
  to $v$.
\end{definition}


\begin{definition}
  \label{def:column:row:indices:1}
  
  Let\footnote{Note the following convention: a row index of
  $\matr{H}$ will be denoted by $j$ and a column index of $\matr{H}$
  will be denoted by $i$.}  $\matr{H} = (h_{ji})$ be the parity-check
  matrix of a binary linear code $\code{C}$. We let $\set{J} \defeq
  \set{J}(\matr{H})$ be the set of row indices of $\matr{H}$ and we
  let $\set{I} \defeq \set{I}(\matr{H})$ be the set of column indices
  of $\matr{H}$, respectively. For each $i \in \set{I}$, we let
  $\set{J}_i \defeq \set{J}_i(\matr{H}) \defeq \big\{ j \in \set{J} \
  | \ h_{ji} {=} 1 \big\}$. For each $j \in \set{J}$, we let
  $\set{I}_j \defeq \set{I}_j(\matr{H}) \defeq \big\{ i \in \set{I} \
  | \ h_{ji} {=} 1 \big\}$. Furthermore, for any $\set{I}' \subseteq
  \set{I}$ and any vector $\vx$ of length $|\set{I}|$, we let
  $\vx_{\set{I}'}$ be the vector that has only the entries of $\vx$
  whose indices are in $\set{I}'$. The Tanner graph~\cite{Tanner:81}
  (or factor graph~\cite{Kschischang:Frey:Loeliger:01}) associated to
  $\matr{H}$ will be called $\graph{T}(\matr{H})$: it consists of bit
  nodes $X_1, \ldots, X_{|\set{I}|}$, (parity-)check nodes $p_1,
  \ldots, p_{|\set{J}|}$, and edges between the two types of
  nodes. More precisely, bit node $i$ and check node $j$ are connected
  if and only if $h_{ji} = 1$. The degree of bit node $i$ is the
  number of adjacent check nodes in $\graph{T}(\matr{H})$ and is
  therefore equal to $|J_i(\matr{H})|$. The degree of check node $j$
  is the number of adjacent bit nodes in $\graph{T}(\matr{H})$ and is
  therefore equal to $|I_j(\matr{H})|$. We say that a vector $\vx \in
  \GF{2}^{|\set{I}|}$ is a configuration of the Tanner graph
  $\graph{T}(\matr{H})$ and we call $\vx \in \GF{2}^{|\set{I}|}$ a
  valid configuration if all the checks are fulfilled, i.e. $\sum_{i
  \in \set{I}} h_{ji} x_i = \sum_{i \in \set{I}_j} x_i = 0 \text{ (in
  $\GF{2}$)}$ for all $j \in \set{J}$. Obviously, the set of all valid
  configurations forms the linear binary code $\code{C}$.
\end{definition}


\begin{definition}
  A binary linear code $\code{C}$ defined by a parity-check matrix
  $\matr{H}$ is called a cycle code if the associated Tanner graph
  $T(H)$ is $2$-regular in the bit nodes, i.e.~all bit nodes have
  degree $2$. This is equivalent to the condition that for all $i \in
  \set{I}(\matr{H})$ we have $|J_i(\matr{H})| = 2$. Such codes were
  studied e.g.~in~\cite{Hakimi:Bredeson:68:1}.
\end{definition}


\begin{figure}
  \begin{center}
    \epsfig{figure=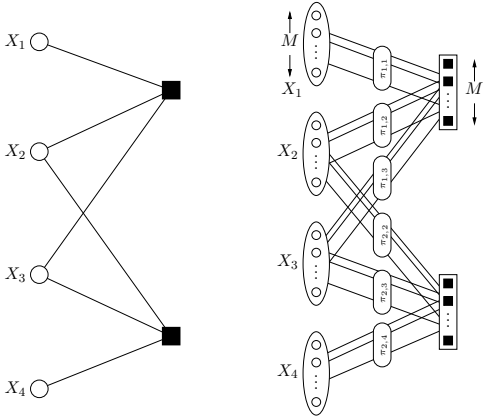,width=2.5in}
  \end{center}
  \caption{ (Code A) Left: Tanner graph $T(\matr{H})$ of the simple
           binary linear code in Ex.~\ref{ex:simple:code:1}. Right:
           Tanner graph of an example of an $M$-cover of
           $T(\matr{H})$.}
  \label{fig:simple:code:1}
\end{figure}


\begin{example}[Code A]
  \label{ex:simple:code:1}

  Let $\code{C}$ be a binary $[4,2]$ code with parity-check matrix
  \begin{align*}
    \matr{H}
      &\defeq
         \begin{pmatrix}
           1 & 1 & 1 & 0 \\
           0 & 1 & 1 & 1
         \end{pmatrix}.
  \end{align*}
  Obviously, $\code{C} = \big\{ (0,0,0,0), (0,1,1,0), (1,0,1,1), (1,1,0,1)
  \big\}$, $\set{J} = \{ 1, 2 \}$, $\set{J}_1 = \{ 1 \}$, $\set{J}_2 = \{ 1, 2
  \}$, $\set{J}_3 = \{ 1, 2 \}$, $\set{J}_4 = \{ 2 \}$, $\set{I} = \{ 1, 2, 3,
  4 \}$, $\set{I}_1 = \{ 1, 2, 3 \}$, and $\set{I}_2 = \{ 2, 3, 4 \}$. The
  Tanner graph $\graph{T}(\matr{H})$ that is associated to $\matr{H}$ is shown
  in Fig.~\ref{fig:simple:code:1} (left); it can easily be seen that this
  is \emph{not} a cycle code.
\end{example}


\begin{example}[Code B]
  \label{ex:simple:code:2}
 
  {\sloppy

  Let $\code{C}$ be a binary $[7,2]$ code with parity-check matrix
  \begin{align*}
    \matr{H}
      &\defeq
         \begin{pmatrix}
           1 & 1 & 0 & 0 & 0 & 0 & 0 \\
           0 & 1 & 1 & 1 & 0 & 0 & 0 \\
           1 & 0 & 1 & 0 & 0 & 0 & 0 \\
           0 & 0 & 0 & 1 & 1 & 0 & 1 \\
           0 & 0 & 0 & 0 & 1 & 1 & 0 \\
           0 & 0 & 0 & 0 & 0 & 1 & 1
         \end{pmatrix}.
  \end{align*}
  Obviously, $\code{C} = \big\{ (0,0,0,0,0,0,0), (1,1,1,0,0,0,0), (0,0,0,0,$
  $1,1,1), (1,1,1,0,1,1,1) \big\}$.\footnote{Note that the rank of $\matr{H}$
  is $5$ and not $6$: therefore the dimension of $\code{C}$ is $2$ and not
  $1$.} The Tanner graph $T(\matr{H})$ of $\code{C}$ is shown in
  Fig.~\ref{fig:simple:code:2} (left). As can easily be seen, all bit nodes
  have degree $2$ and so the code $\code{C}$ is a cycle code. From the Tanner
  graph $T(\matr{H})$ we can derive another graph $N(\matr{H})$ in the
  following way: replace each (degree-$2$) bit node and its adjacent edges by
  a single edge and label the new edge according to the labeling of the bit
  node in the Tanner graph.\footnote{We gave the label $N(\matr{H})$ because
  such a graph is also known as normal graph or Forney-style factor
  graph~\cite{Forney:01:1}.} For code $\code{C}$ we obtain the graph
  $N(\matr{H})$ shown in Fig.~\ref{fig:simple:code:2} (right). From this graph
  the notion of ``cycle code'' becomes clear: every codeword (i.e.~every valid
  configuration) corresponds to a simple cycle or a symmetric difference set of
  simple cycles in the normal graph. This will be made more precise in
  Sec.~\ref{sec:zeta:functions:1}.

  }
\end{example}


\begin{figure}[t]
  \begin{center}
    \epsfig{figure=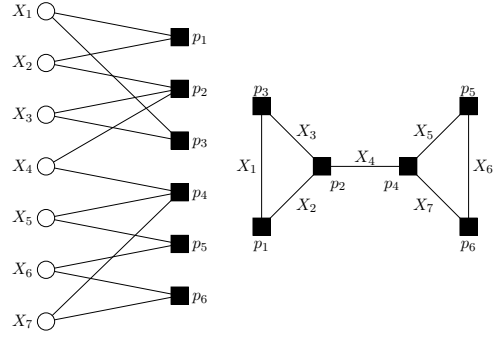, width=2.5in}
  \end{center}
  \caption{ (Code B) Left: Tanner graph $T(\matr{H})$ of the cycle code
           $\code{C}$ in Ex.~\ref{ex:simple:code:2}. Right: Normal graph
           $N(\matr{H})$ of the cycle code $\code{C}$ in
           Ex.~\ref{ex:simple:code:2}.}
  \label{fig:simple:code:2}
\end{figure}


\begin{figure}
  \begin{center}
    \epsfig{figure=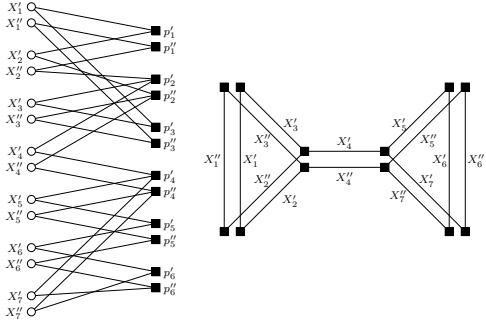,width=2.5in}
  \end{center}
  \caption{(Code B) Left: A double cover of the Tanner graph
           $T(\matr{H})$ in Fig.~\ref{fig:simple:code:2}
           (left). Right: The corresponding double cover of the normal
           graph $N(\matr{H})$ in Fig.~\ref{fig:simple:code:2}
           (right).}
  \label{fig:simple:code:2:cover:1}
\end{figure}


\itwsection{The Fundamental Cone}

\label{sec:fundamental:cone}


The following definition introduces the  graph theoretic notion of a
``graph cover''.


\begin{definition}
  \cite{Massey:77:1, Stark:Terras:96:1} An {\em unramified, finite cover}, or,
  simply, a {\em cover} of a graph $X$ is a graph $Y$ along with a surjective
  map $\pi:Y \to X$ which is a graph homomorphism, i.e., which takes adjacent
  vertices of $Y$ to adjacent vertices of $X$, such that for each vertex $x$
  of $X$ and each $y \in \pi^{-1}(x)$, the neighborhood $\del(y)$ of $y$ is
  mapped bijectively to $\del(x)$.  For a positive integer $M$, an {\em
  $M$-cover} of $X$ is an unramified finite cover $\pi:Y \to X$ such that for
  each vertex $x$ of $X$, $\pi^{-1}(x)$ contains exactly $M$ vertices of $Y$.
\end{definition}


\begin{example}[Code A]
  \label{ex:simple:code:1:cover:1}

  We continue with Code A defined in Ex.~\ref{ex:simple:code:1}. Let
  $\graph{T} \defeq \graph{T}(\matr{H})$ be the Tanner graph corresponding to
  $\matr{H}$. An $M$-fold cover $\graph{\widetilde T}$ (as shown in
  Fig.~\ref{fig:simple:code:1} (right)) of $\graph{T}$ is specified by
  defining the permutations $\pi_{1,1}$, $\pi_{1,2}$, $\pi_{1,3}$
  (corresponding to the first row of $\matr{H}$) and the permutations
  $\pi_{2,2}$, $\pi_{2,3}$, $\pi_{2,4}$ (corresponding to the second row of
  $\matr{H}$).

  The parity-check matrix $\matr{\widetilde H}$ associated to one
  possible $3$-fold cover Tanner graph $\graph{\widetilde T}$ looks
  like
  \begin{align*}
    \matr{\widetilde H}
      &\defeq
         \left(
           \begin{array}{ccc|ccc|ccc|ccc}
             0 & 1 & 0  &  1 & 0 & 0  &  0 & 1 & 0  &  0 & 0 & 0 \\
             0 & 0 & 1  &  0 & 1 & 0  &  0 & 0 & 1  &  0 & 0 & 0 \\
             1 & 0 & 0  &  0 & 0 & 1  &  1 & 0 & 0  &  0 & 0 & 0 \\
             \hline
             0 & 0 & 0  &  0 & 0 & 1  &  0 & 0 & 1  &  1 & 0 & 0 \\
             0 & 0 & 0  &  1 & 0 & 0  &  1 & 0 & 0  &  0 & 1 & 0 \\
             0 & 0 & 0  &  0 & 1 & 0  &  0 & 1 & 0  &  0 & 0 & 1
           \end{array}
         \right) \\
       &= \begin{pmatrix}
           \matrunity_2 & \matrunity_0 & \matrunity_2 & \matr{0} \\
           \matr{0}     & \matrunity_1 & \matrunity_1 & \matrunity_0
         \end{pmatrix},
  \end{align*}
  where $\matrunity_s$ is a $3 \times 3$ identity matrix, cyclically
  shifted to the left by $s$ positions. This parity-check matrix
  defines a code $\code{\widetilde{C}}$: an example of a codeword of
  $\code{\widetilde{C}}$ is $\tvc = (1,1,0, \ 0,1,1, \ 0,1,1, \
  0,0,0)$.
\end{example}


\begin{figure}
  \begin{center}
    \epsfig{figure=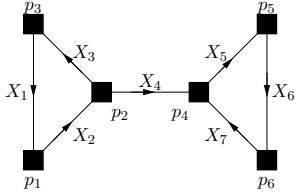,width=1.5in}
  \end{center}
  \caption{(Code B) A directed normal graph of the normal graph
           $N(\matr{H})$ in Fig.~\ref{fig:simple:code:2} (left).}
  \label{fig:simple:code:2:directed:1}
\end{figure}


Other examples of a graph cover are shown in
Fig.~\ref{fig:simple:code:2:cover:1}: the left-hand side shows a double cover
Tanner graph of the Tanner graph in Fig.~\ref{fig:simple:code:2} (left) and
the right-hand side shows the corresponding double cover normal graph of the
normal graph in Fig.~\ref{fig:simple:code:2} (right). The following remark
formalizes Ex.~\ref{ex:simple:code:1:cover:1}.


\begin{remark}\label{rmk:temp:1}
  Let $\code{C}$ be a binary code with parity-check matrix $\matr{H}$
  and Tanner graph $\graph{T} \defeq \graph{T}(\matr{H})$. Let
  $\set{J} \defeq \set{J}(\matr{H})$ and $\set{I}_j \defeq
  \set{I}_j(\matr{H})$. For a positive integer $M$, let
  $\graph{\widetilde T}$ be an arbitrary $M$-fold cover of $\graph{T}$
  and let $\code{\widetilde C}$ be the binary code described by
  $\graph{\widetilde T}$. Knowing the graph $\graph{T}$, the graph
  $\graph{\widetilde T}$ is completely specified by defining for all
  $j \in \set{J}$ and all $i \in \set{I}_j$ the permutations
  $\pi_{j,i}$ that map $[M] \defeq \{ 1, \ldots, M \}$ onto
  itself. The meaning of $\pi_{j,i}(m)$, $m \in [M]$ is the following:
  the $m^{\text{th}}$ copy of the check node $j$ is connected to the
  $\pi_{j,i}(m)^{\text{th}}$ copy of the $i^{\text{th}}$ bit.  It
  follows that $\tvc \in \code{\widetilde C}$ if and only if
  \begin{align*}
    \sum_{i \in \set{I}_j}
      \widetilde c_{i,\pi_{j,i}(m)}
      &= 0 \quad \text{(in $\GF{2}$)}
             \label{eq:remark:parity:check:equality:cover:code:1}
  \end{align*}
  for all $j \in \set{J}$ and all $m \in [M]$. The parity-check matrix
  $\matr{\widetilde H}$ that expresses this fact can be defined as
  follows. Let the entries of $\matr{\widetilde H}$ be indexed by $(j,m) \in
  \set{J} \times [M]$ and $(i,m') \in \set{I} \times [M]$.  Then
  \begin{align*}
    h_{(j,m),(i,m')}
      &\defeq
         \begin{cases}
           1 & \text{if $i \in \set{I}_j$ and $m' = \pi_{j,i}(m)$} \\
           0 & \text{otherwise}.
         \end{cases}
  \end{align*}
\end{remark}


\begin{definition}
  \label{def:pseudo:codeword:1}
  
  \cite{Koetter:Vontobel:03:1} Let $\code{C}$ be a binary linear (base) code
  with parity-check matrix $\matr{H}$ and let $\graph{T} \defeq
  \graph{T}(\matr{H})$ be the corresponding Tanner graph. For any positive
  integer $M$, let $\graph{\widetilde T}$ be an $M$-fold cover of $\graph{T}$
  and let $\code{\widetilde C}$ be the binary code described by
  $\graph{\widetilde T}$. We will denote a codeword of $\code{\widetilde C}$
  by $\tvc$, where the $(i,m)$'s component of $\tvc$, i.e.~$\widetilde
  c_{i,m}$, denotes the value of the $m^{\text{th}}$ copy of the
  $i^{\text{th}}$ bit.
  
  The {\em pseudo-codeword} associated to $\tvc$ is the rational
  vector $\vomega(\tvc) \defeq \big( \omega_1(\tvc), \omega_2(\tvc), \ldots,
  \omega_n(\tvc) \big)$ with
  \begin{align*}
    \omega_i(\tvc)
      &\defeq
         \frac{1}{M}
           \sum_{m \in [M]}
             \widetilde c_{i,m},
  \end{align*}
  where the sum is taken in $\mathbb{R}$ (not in $\GF{2}$). We call the vector
  $M \cdot \vomega(\tvc)$ the \emph{unscaled} pseudo-codeword
  associated with $\tvc$. In fact, any multiple (by a positive scalar) of
  $\vomega(\tvc)$ will be called a pseudo-codeword associated with $\tvc$.
\end{definition}

Note that any codeword is also a pseudo-codeword.


\begin{remark}
  Notice that a pseudo-codeword, as defined in
  Def.~\ref{def:pseudo:codeword:1}, has length $|\set{I}(\matr{H})|$, the same
  as the length of any codeword, whereas a codeword like $\tvc \in
  \code{\widetilde{C}}$ has length $M \cdot |\set{I}(\matr{H})|$, where $M$ is
  the degree of the corresponding cover Tanner graph.

\end{remark}


\begin{example}[Code A]
  \label{ex:simple:code:1:pseudo:codeword:1}

  We continue with Code A defined in Ex.~\ref{ex:simple:code:1}. We saw that
  $\tvc = (1,1,0, \ 0,1,1, \ 0,1,1, \ 0,0,0)$ was a codeword of the code
  $\code{\widetilde{C}}$. Applying Def.~\ref{def:pseudo:codeword:1} we see
  that the corresponding pseudo-codeword is $\vomega(\tvc) = (\frac{2}{3},
  \frac{2}{3}, \frac{2}{3}, 0)$ and that the corresponding unscaled
  pseudo-codeword is $3 \cdot \vomega(\tvc) = (2,2,2,0)$. Note that
  $\vomega(\tvc)$ cannot be written as a convex combination of the codewords
  in $\code{C}$.
\end{example}


The influence of a pseudo-codeword on the decoding behavior under iterative
decoding can be measured by its pseudo-weight which is a function of the
pseudo-codeword and the channel used (see~\cite{Koetter:Vontobel:03:1} and
references therein). An important property of the pseudo-weight is its scaling
invariance, i.e.~scaling a pseudo-codeword by a positive scalar leaves
its pseudo-weight unchanged.



%




The fundamental cone that is given in the following definition will
be, along with the zeta functions of a graph, a main object of interest in
this paper.


\begin{definition}
  \label{def:fundamental:cone:1}

  \cite{Koetter:Vontobel:03:1, Feldman:Karger:Wainwright:03:2} Let
  $\code{C}$ be an arbitrary binary linear code and let $\matr{H}$ be
  its parity-check matrix. We define the {\em fundamental cone}
  $\fch{K}{H}$ of $\matr{H}$ to be the set of vectors $\vomega \in
  \R^n$ that satisfy
  \begin{alignat*}{2}
    \forall i \in \set{I}:&&
        \quad
    \omega_i
      &\geq 0, \\
    \forall j \in \set{J}, \
      \forall i \in \set{I}_j:&&
        \quad
    \sum_{i' \in \set{I}_j \setminus \{ i \}}
      \omega_{i'}
      &\geq
        \omega_{i},
  \end{alignat*}
  where $\set{J} \defeq \set{J}(\matr{H})$, $\set{I} \defeq
  \set{I}(\matr{H})$, $\set{I}_j \defeq \set{I}_j(\matr{H})$.
\end{definition}


\begin{example}[Code A]
  \label{ex:simple:code:1:fundamental:cone:1}

  We continue with Code A defined in Ex.~\ref{ex:simple:code:1}. The
  fundamental cone $\fch{K}{H}$ is the set
  \begin{align*}
    \fch{K}{H}
      = \{
           (\omega_1,\omega_2,&\omega_3,\omega_4) \in \R^4
           \big|\,
                 \omega_1 \geq 0,
                 \omega_2 \geq 0,
                 \omega_3 \geq 0,
                 \omega_4 \geq 0, \\
                 &- \omega_1 + \omega_2 + \omega_3 \geq 0, \ \ \
                 - \omega_2 + \omega_3 + \omega_4 \geq 0, \\
                 &+ \omega_1 - \omega_2 + \omega_3 \geq 0, \ \ \
                 + \omega_2 - \omega_3 + \omega_4 \geq 0, \\
                 &+ \omega_1 + \omega_2 - \omega_3 \geq 0, \ \ \
                 + \omega_2 + \omega_3 - \omega_4 \geq 0
         \}.  
  \end{align*}
\end{example}




The next two lemmas
establish that there is
a very tight connection between the fundamental cone of a code and codewords
that live in finite covers. More specifically, in one direction we prove that
the pseudo-codeword associated to any codeword in a cover of a Tanner graph
must lie within the fundamental polytope. In the other direction we prove that
to a given vector in the fundamental polytope we can find a cover with a
codeword in it whose (suitably scaled) pseudo-codeword is arbitrarily close to
the given vector.  



\begin{lemma}
  \label{lemma:from:cover:codeword:to:pseudo:codeword:1}

  \cite{Koetter:Vontobel:03:1} Let $\code{C}$ be a binary linear code with
  parity-check matrix $\matr{H}$ and Tanner graph $\graph{T} =
  \graph{T}(\matr{H})$. For a positive integer $M$, let $\graph{\widetilde T}$
  be an arbitrary $M$-fold cover of $\graph{T}$ and let $\code{\widetilde C}$
  be the binary code described by $\graph{\widetilde T}$. If $\tvc \in
  \code{\widetilde C}$ then $\vomega(\tvc) \in \fch{K}{H}$.
\end{lemma}


\begin{lemma}
  \label{lemma:from:pseudo:codeword:to:cover:codeword:1}

  \cite{Koetter:Vontobel:03:1} Let $\code{C}$ be a binary linear code with
  parity-check matrix $\matr{H}$ and Tanner graph $\graph{T} =
  \graph{T}(\matr{H})$. Let the vector $\vomega' \in \R^n$ satisfy $\vomega'
  \in \fch{K}{H}$. Then for any $\varepsilon > 0$, there is a positive integer
  $M$ such that there is a codeword $\tvc$ in a code $\code{\widetilde{C}}$
  defined by an $M$-fold cover $\graph{\widetilde T}$ of $\graph{T}$ such that
  $||\alpha \vomega(\tvc) - \vomega'||_2 < \varepsilon$ for some $\alpha >
  0$. 
\end{lemma}


Putting Lemmas~\ref{lemma:from:cover:codeword:to:pseudo:codeword:1}
and \ref{lemma:from:pseudo:codeword:to:cover:codeword:1} together, we
have: 


\begin{theorem} Let $C$ be a binary linear code with parity-check matrix $H$
  and fundamental cone $\fch{K}{H}$.  Then the lines through the
  pseudo-codewords for $C$ are dense in $\fch{K}{H}$. \qed
\end{theorem}


Moreover, we have


\begin{theorem} 
  The point $\vomega = (\omega_1, \dots, \omega_{|\set{I}|}) \in K(H) \cap
  \Z^n$ is an unscaled pseudo-codeword if and only if $\sum_{i \in \set{I}}
  h_{ji} \omega_i = 0 \text{ (in $\GF{2}$)}$ for each $j \in \set{J}$.
\end{theorem}


\begin{proof}This follows from 
  Lemma~\ref{lemma:from:pseudo:codeword:to:cover:codeword:1} and
  Corollary~\ref{cor:to:pseudos:in:cycle}.
\end{proof}


\itwsection{Zeta Functions of Graphs}

\label{sec:zeta:functions:1}


Before we can talk about zeta functions of graphs we need to say exactly what
we mean by a cycle in a graph.

\begin{definition}
  Let $X$ be an undirected graph as in Def.~\ref{def:undirected:graph:1}. A
  sequence $(e_{i_1}, \dots, e_{i_k})$ of edges of $X$ is a {\em cycle} on $X$
  if the edges $e_{i_j}$ can be directed so that $e_{i_s}$ terminates where
  $e_{i_{s+1}}$ begins for $1 \leq s \leq k-1$ and $e_{i_k}$ terminates where
  $e_{i_1}$ begins. The {\em characteristic vector} of the cycle $(e_{i_1},
  \dots, e_{i_k})$ on $X$ is the binary vector of length $n$ whose
  $t^\text{th}$ coordinate is 1 if and only if $e_t$ appears as some
  $e_{i_j}$. If 
  the cycle does not cross itself, i.e., if each vertex of $X$ is
  involved in at most two of the edges $e_{i_1}$, \dots, $e_{i_k}$,
  then we say the cycle is {\em simple}.
\end{definition}


This definition relates as follows to the cycle codes introduced in
Sec.~\ref{sec:binary:linear:codes:1}:


\begin{lemma}
  \label{lemma:cycle:code:1}

  Let $N \defeq N(\matr{H})$ be the normal graph of a binary cycle
  code $\code{C}$ with parity-check matrix $\matr{H}$. The
  characteristic vector of any simple cycle in $N$ is a valid
  configuration of $N$, i.e.~it is a codeword of $\code{C}$. Moreover,
  the symmetric difference of the characteristic vector of simple
  cycles in $N$ is also a valid configuration of $N$, i.e.~it is a
  codeword of $\code{C}$. On the other hand, to any codeword in
  $\code{C}$ corresponds the symmetric difference of simple cycles in
  $N$.
\end{lemma}


\begin{proof}
  This follows from Euler's Theorem~\cite[Th.~1.2.26]{West:96:1}.
\end{proof}


The code $\code{C}$ in Lemma~\ref{lemma:cycle:code:1} can also be seen
as spanned by the characteristic vectors of the simple cycles of $N$.
The length of $\code{C}$ equals $n(N)$, the number of edges in
$N$. Further, the minimum Hamming distance $\dmin$ of $\code{C}$ is
the length of the shortest cycle in $N$, i.e., the girth of $N$.
Also, the dimension of $\code{C}$ is the number of independent cycles
in $N$, i.e., the rank of the fundamental group of the underlying
topological space of $N$, i.e., $|E(N)| - |V(N)| + 1 = 1-\chi(N)$,
where $\chi(N)$ is the Euler characteristic of $N$.


Let us turn back to graph-theoretic notions: the next important step
is to introduce a special class of cycles called ``primitive,
backtrackless and tailless cycles''.


\begin{definition}\label{primitive:backtrackless:tailless:equivalence}
  Let $\Gamma = (e_{i_1}, \dots, e_{i_k})$ be a cycle in a graph
  $X$. We say $\Gamma$ is {\em backtrackless} if for no $s$ do we have
  $e_{i_s} = e_{i_{s+1}}$.  We say $\Gamma$ is {\em tailless} if
  $e_{i_1} \neq e_{i_k}$.  We say $\Gamma$ is {\em primitive} if there
  is no cycle $\Theta$ on $X$ such that $\Gamma = \Theta^r$ with $r
  \geq 2$, i.e., such that $\Gamma$ is obtained by following $\Theta$
  a total of $r$ times.  We say that the cycle $\Psi = (e_{j_1},
  \dots, e_{j_k})$ is {\em equivalent} to $\Gamma$ if there is some
  integer $t$ such that $e_{j_s} = e_{j_{(s+t)\ \operatorname{mod}\
  k}}$ for all $s$.
\end{definition}


It is easy to check that any simple cycle is a primitive,
backtrackless and tailless cycle and that the notion of equivalence
given in Def.~\ref{primitive:backtrackless:tailless:equivalence}
defines an equivalence relation on primitive, backtrackless, tailless
cycles.


\begin{example}[Code B]
  \label{ex:simple:code:2:cycles:1}

  Let us return to Code B defined in Ex.~\ref{ex:simple:code:2} and its
  normal graph shown in Fig.~\ref{fig:simple:code:2} (right); the edge with
  variable label $X_i$ will be called $e_i$. We see that the edge-sequences
  $(e_1,e_2,e_3)$ and $(e_5,e_6,e_7)$ are simple cycles: they correspond to
  the codewords $(1,1,1,0,0,0,0)$ and $(0,0,0,0,1,1,1)$, respectively, in
  $\code{C}$.

  In contrast to these two cycles, the cycles
  \begin{align*}
    \Gamma_1
      &= (e_1, e_2, e_4, e_5, e_6, e_7, e_4, e_3) \\
    \Gamma_2
      &= (e_3, e_4, e_7, e_6, e_5, e_4, e_2, e_1) \\
    \Gamma_3
      &= (e_1, e_2, e_4, e_5, e_6, e_7, e_5, e_6, e_7, e_4, e_3)
  \end{align*}
  are \emph{not} simple cycles; but they are inequivalent, backtrackless,
  tailless, primitive cycles. Indeed, we can obtain infinitely many
  inequivalent, backtrackless, tailless, primitive cycles on $N(\matr{H})$ by,
  for example, following the path $(e_1,e_2,e_4)$, then arbitrarily many
  copies of the loop $(e_5,e_6,e_7)$, and then $(e_4,e_3)$.
\end{example}


The \emph{edge zeta function} of a graph is a way to enumerate all
inequivalent, primitive, backtrackless cycles and combinations thereof.


\begin{definition}
  \label{def:edge:zeta:function:1}

  \cite{Hashimoto:89:1, Stark:Terras:96:1} Let $\Gamma$ be a path in a graph
  $X$ with edge-set $E$; write $\Gamma = (e_{i_1}, \dots, e_{i_k})$ to
  indicate that $\Gamma$ begins with the edge $e_{i_1}$ and ends with the edge
  $e_{i_k}$. The {\em monomial of $\Gamma$} is given by $g(\Gamma)
  \defeq u_{i_1}\cdots u_{i_k}$,
  where the $u_i$'s are indeterminates. The {\em edge zeta function} of
  $X$ is defined to be the power series $\zeta_X(u_1, \dots, u_n) \in \Z[[u_1,
  \dots, u_n]]$ given by
\vspace{-1mm}
  \begin{align*}
    \zeta_X(u_1, \dots, u_n)
      &= \prod_{[\Gamma] \in A(X)}
           \big(
             1 - g(\Gamma)
           \big)^{-1},
  \end{align*}
  where $A(X)$ is the collection of equivalence classes of backtrackless,
  tailless, primitive cycles in $X$.
\end{definition}


As Ex.~\ref{ex:simple:code:2:cycles:1} shows, the product in the definition of
the edge zeta function is, in general, infinite. However, it is true that the
edge zeta function is a rational function.  To see this, we first need a few
more definitions.


\begin{definition}
  \label{def:directed:graph:1}

  \cite{Stark:Terras:96:1} Let $X = (V(X),E(X))$ be an undirected graph with
  edge set $E(X) = \{ e_1, \ldots, e_n \}$. A {\em directed graph} $\dirX =
  (\dirV(\dirX),\dirE(\dirX))$ derived from $X$ is a graph with vertex set
  $\dirV(\dirX) \defeq V(X)$ and edge set $\dirE(\dirX) \defeq \{f_1, \dots,
  f_{2n}\}$, where the (directed) edges $f_i$ and $f_{n+i}$ both correspond to
  the same edge $e_i \in E(X)$ but have opposite directions.
\end{definition}


\begin{definition} \cite{Stark:Terras:96:1} Let $\dirX =
  (\dirV(\dirX),\dirE(\dirX))$ be a directed graph as defined in
  Def.~\ref{def:directed:graph:1}. The {\em directed edge matrix} of $\dirX$
  is the matrix $M(\dirX) = (m_{ij})$ where
  \begin{align*} 
    m_{ij}
      &= \begin{cases}
           1, & \text{if $f_i$ feeds into $f_j$ 
                      to form a backtrackless path} \\
           0, &\text{otherwise.}
         \end{cases}
  \end{align*}
\end{definition}


\begin{example}[Code B]
  \label{ex:simple:code:2:directed:graph:1}

  Let us continue with Code B defined in Ex.~\ref{ex:simple:code:2}. The
  normal graph $N = N(\matr{H})$ of the code is shown in
  Fig.~\ref{fig:simple:code:2} (right); the edge with variable label $X_i$
  will be called $e_i$. The directed edges $f_1$ to $f_{14}$ of a directed
  version $\dirN$ of $N$ are chosen such that the edges $f_1$ to $f_7$ are as
  shown in Fig.~\ref{fig:simple:code:2:directed:1}.  Implicitly this figure
  also defines the edges $f_8$ to $f_{14}$; e.g., $f_{11}$ is the same as
  $f_4$ but directed from right to left. The directed edge matrix $M \defeq
  M(\dirN)$ of $\dirN$ is then the matrix { \small
  \begin{align*}
    M
      &= \left[
           \begin{array}{ccccccc|ccccccc}
             0 & 1 & 0 & 0 & 0 & 0 & 0 & 0 & 0 & 0 & 0 & 0 & 0 & 0 \\
             0 & 0 & 1 & 1 & 0 & 0 & 0 & 0 & 0 & 0 & 0 & 0 & 0 & 0 \\
             1 & 0 & 0 & 0 & 0 & 0 & 0 & 0 & 0 & 0 & 0 & 0 & 0 & 0 \\
             0 & 0 & 0 & 0 & 1 & 0 & 0 & 0 & 0 & 0 & 0 & 0 & 0 & 1 \\
             0 & 0 & 0 & 0 & 0 & 1 & 0 & 0 & 0 & 0 & 0 & 0 & 0 & 0 \\ 
             0 & 0 & 0 & 0 & 0 & 0 & 1 & 0 & 0 & 0 & 0 & 0 & 0 & 0 \\
             0 & 0 & 0 & 0 & 1 & 0 & 0 & 0 & 0 & 0 & 1 & 0 & 0 & 0 \\ 
             \hline
             0 & 0 & 0 & 0 & 0 & 0 & 0 & 0 & 0 & 1 & 0 & 0 & 0 & 0 \\ 
             0 & 0 & 0 & 0 & 0 & 0 & 0 & 1 & 0 & 0 & 0 & 0 & 0 & 0 \\
             0 & 0 & 0 & 1 & 0 & 0 & 0 & 0 & 1 & 0 & 0 & 0 & 0 & 0 \\ 
             0 & 0 & 1 & 0 & 0 & 0 & 0 & 0 & 1 & 0 & 0 & 0 & 0 & 0 \\ 
             0 & 0 & 0 & 0 & 0 & 0 & 0 & 0 & 0 & 0 & 1 & 0 & 0 & 1 \\ 
             0 & 0 & 0 & 0 & 0 & 0 & 0 & 0 & 0 & 0 & 0 & 1 & 0 & 0 \\ 
             0 & 0 & 0 & 0 & 0 & 0 & 0 & 0 & 0 & 0 & 0 & 0 & 1 & 0
           \end{array}
         \right].
  \end{align*}
  }%
\end{example}


With these definitions, Stark and Terras \cite{Stark:Terras:96:1} prove:


\begin{theorem}
  \cite{Stark:Terras:96:1} The edge zeta function $\zeta_X(u_1, \dots, u_n)$
  is a rational function. More precisely, for any directed graph $\dirX$ of
  $X$, we have
  \begin{align*}
    \zeta_X(u_1, \dots, u_n)^{-1}
      &= \det(I-UM(\dirX)) = \det(I-M(\dirX)U)
  \end{align*}
  where $I$ is the identity matrix of size $2n$ and $U =
  \mathop{\text{diag}}(u_1, \dots, u_n, u_1, \dots, u_n)$ is a
  diagonal matrix of indeterminants.
\end{theorem}


\begin{example}[Code B]
  \label{ex:simple:code:2:edge:zeta:1}

  Let us continue with Code B defined in Ex.~\ref{ex:simple:code:2} and its
  normal graph $N \defeq N(\matr{H})$. By the above theorem and using $\dirN$
  from Ex.~\ref{ex:simple:code:2:directed:graph:1}, the edge zeta function
  $\zeta_N$ of our graph $N$ satisfies
  \begin{align*}
  \zeta_N&(u_1, \dots, u_7)^{-1}
     = \det(I_{14}-U M)
     = \det(I_{14}-M U)\\
    &= 1 - 2 u_1 u_2 u_3 + u_1^2 u_2^2 u_3^2 - 2 u_5 u_6 u_7
         + 4 u_1 u_2 u_3 u_5 u_6 u_7 \\
    & \phantom{spa} - 2 u_1^2 u_2^2 u_3^2 u_5 u_6 u_7
                    - 4 u_1 u_2 u_3 u_4^2 u_5 u_6 u_7 \\
    & \phantom{spa} + 4 u_1^2 u_2^2 u_3^2 u_4^2 u_5 u_6 u_7
                    + u_5^2 u_6^2 u_7^2
                    - 2 u_1 u_2 u_3 u_5^2 u_6^2 u_7^2 \\
    & \phantom{spa} + u_1^2 u_2^2 u_3^2 u_5^2 u_6^2 u_7^2
                    + 4 u_1 u_2 u_3 u_4^2 u_5^2 u_6^2 u_7^2 \\
    & \phantom{spa} - 4 u_1^2 u_2^2 u_3^2 u_4^2 u_5^2 u_6^2 u_7^2.
  \end{align*}
  Expanding out the Taylor series, we get the first several terms of
  $\zeta_N$:
  \begin{align*}
    \zeta_N&(u_1, \dots, u_7)
       = 1 + 2 u_1 u_2 u_3 + 3 u_1^2 u_2^2 u_3^2
           + 2 u_5 u_6 u_7 \\ 
      &\phantom{spa} + 4 u_1 u_2 u_3 u_5 u_6 u_7
                     + 6 u_1^2 u_2^2 u_3^2 u_5 u_6 u_7 \\ 
      &\phantom{spa} + 4 u_1 u_2 u_3 u_4^2 u_5 u_6 u_7
                     + 12 u_1^2 u_2^2 u_3^2 u_4^2 u_5 u_6 u_7 \\ 
      &\phantom{spa} + 3 u_5^2 u_6^2 u_7^2
                     + 6 u_1 u_2 u_3 u_5^2 u_6^2 u_7^2
                     + 9 u_1^2 u_2^2 u_3^2 u_5^2 u_6^2 u_7^2 \\ 
      &\phantom{spa} + 12 u_1 u_2 u_3 u_4^2 u_5^2 u_6^2 u_7^2
                     + 36 u_1^2 u_2^2 u_3^2 u_4^2 u_5^2 u_6^2 u_7^2
                     + \cdots.
  \end{align*}
\end{example}


\itwsection{Relating the Fundamental Cone 
            and the Zeta Function of a Cycle Code}

\label{sec:connection:fundamental:cone:zeta:function:1}


The results of this chapter are based on the simple observations made in the
following example.


\begin{example}[Code B]
  \label{ex:simple:code:2:connection:fundamental:cone:zeta:function:1}

  Let us continue with Code B defined in Ex.~\ref{ex:simple:code:2} and its
  Tanner graph $T \defeq T(\matr{H})$ as shown in Fig.~\ref{fig:simple:code:2}
  (left). We saw that any codeword corresponds one-to-one to a valid
  configuration in $T$.

  Consider now a double cover $\widetilde{T}$ of $T$ as shown in
  Fig.~\ref{fig:simple:code:2:cover:1} (left): the set of all valid
  configurations of $\widetilde{T}$ defines a code
  $\code{\widetilde{C}}$. Because of the properties of graph covers,
  the code $\code{\widetilde{C}}$ is again a cycle code and in the
  same manner as in Ex.~\ref{ex:simple:code:2} we deduce its normal
  graph $\widetilde{N}$. It is not hard to see that $\widetilde{N}$
  shown in Fig.~\ref{fig:simple:code:2:cover:1} (right) is a double
  cover of the normal graph $N \defeq N(\matr{H})$ shown in
  Fig.~\ref{fig:simple:code:2} (right).


  Just as the codewords of $\code{C}$ correspond bijectively to the
  vectors in the span of the characteristic vectors of the simple
  cycles in $N$, the codewords of $\code{\widetilde{C}}$ correspond
  bijectively to the vectors in the span of the characteristic vectors
  of the simple cycles in $\widetilde{N}$.

  An example of simple cycle in $\widetilde{N}$ is the
  edge-sequence\footnote{The edge with variable label $X'_i$ ($X''_i$)
  will be called $e'_i$ ($e''_i$).}
  \begin{align*}
    \widetilde{\Gamma}
      &= (e'_1, e'_2, e''_4, e''_5, e''_6, e'_7, e'_4, e'_3).
  \end{align*}
  After mapping it down to $N$ it reads
  \begin{align*}
    \pi(\widetilde{\Gamma})
      &= (e_1, e_2, e_4, e_5, e_6, e_7, e_4, e_3),
  \end{align*}
  which is a backtrackless and tailless cycle in $N$ which is not simple. Note
  that in general the image of a simple cycle is always backtrackless and
  tailless, but not necessarily simple or primitive. The cycle
  $\widetilde{\Gamma}$ corresponds to a codeword $\tvc$ and the mapped cycle
  $\pi(\widetilde{\Gamma})$ corresponds to the pseudo-codeword
  \begin{align*}
    \vomega(\tvc)
      &= \frac{1}{2}
           \cdot
           (1, 1, 1, 2, 1, 1, 1)
       = \left(
           \frac{1}{2}, \frac{1}{2}, \frac{1}{2}, 1, 
           \frac{1}{2}, \frac{1}{2}, \frac{1}{2}
         \right).
  \end{align*}
\end{example}

With this example we can draw the following important conclusion about cycle
codes (which will be formalized in Th.~\ref{theorem:pseudos:in:cycle:1}):
listing the pseudo-codewords stemming from all the possible finite covers is
equivalent to listing all backtrackless and tailless cycles of the normal
graph and combinations thereof. But listing these cycles (in a certain way) is
exactly what the zeta function of the normal graph essentially does!



\begin{definition}
  The {\em exponent vector} of the monomial $u_1^{p_1}\dots u_n^{p_n}$ is the
  vector $(p_1, \dots, p_n) \in \N_0^n$ of the exponents of the monomial.
\end{definition}




\begin{example}[Code B]
  Continuing with Code B that was defined in Ex.~\ref{ex:simple:code:2} and
  the zeta function $\zeta_N$ of its normal graph $N \defeq N(\matr{H})$
  (cf.~Ex.~\ref{ex:simple:code:2:edge:zeta:1}), we see that the exponent
  vectors of the first several monomials appearing in $\zeta_N$ are
  (0,0,0,0,0,0,0), (1,1,1,0,0,0,0), (2,2,2,0,0,0,0), (0,0,0,0,1,1,1),
  (1,1,1,0,1,1,1), (2,2,2,0,1,1,1), (1,1,1,2,1,1,1), (2,2,2,2,1,1,1),
  (0,0,0,0,2,2,2), (1,1,1,0,2,2,2), (2,2,2,0,2,2,2), (1,1,1,2,2,2,2),
  (2,2,2,2,2,2,2), \dots. Note that most of these lie within the span of
  multiples of codewords in $\code{C}$; for example,
  \begin{align*}
    (1,1,1,0,2,2,2)
      &= (1,1,1,0,0,0,0) + 2(0,0,0,0,1,1,1).
  \end{align*} 
  The exceptions thus far are (1,1,1,2,1,1,1), (2,2,2,2,1,1,1),
  (1,1,1,2,2,2,2) and (2,2,2,2,2,2,2). The first of these exceptions is
  exactly the pseudo-codeword for $C$ given in
  Ex.~\ref{ex:simple:code:2:connection:fundamental:cone:zeta:function:1}, and
  the rest lie within the span of this pseudo-codeword along with multiples of
  codewords.
\end{example}


These observations are made precise in the next theorem.


\begin{theorem}
  \label{theorem:pseudos:in:cycle:1}

  Let $C$ be a cycle code defined by a parity-check matrix $\matr{H}$
  having normal graph $N \defeq N(\matr{H})$, let $n = n(N)$ be the
  number of edges of $N$, and let $\zeta_N(u_1, \dots, u_n)$ be the
  edge zeta function of $N$. Then the monomial $u_1^{p_1} \dots
  u_n^{p_n}$ has nonzero coefficient in $\zeta_N$ if and only if the
  corresponding exponent vector $(p_1, \dots, p_n)$ is an unscaled
  pseudo-codeword for $C$.
\end{theorem}


\begin{proof}[Sketch of proof.]
  By Def.~\ref{def:edge:zeta:function:1}, the monomial $u_1^{p_1}\dots
  u_n^{p_n}$ appears with nonzero coefficient in $\zeta_N$ if and only if
  there are backtrackless, tailless, primitive cycles $\Gamma_1, \dots,
  \Gamma_m$ on $X$ such that
  \begin{align*}
    u_1^{p_1}\dots u_n^{p_n}
      &= g(\Gamma_1)^{q_1} \cdots g(\Gamma_m)^{q_m}
  \end{align*} 
  for some nonnegative integers $q_1, \dots, q_m$.  It is thus enough to prove
  that $\Gamma$ is a backtrackless, tailless cycle on $N$ if and only if
  $\Gamma = \pi(\widetilde \Gamma)$ for some simple cycle $\widetilde \Gamma$
  on some (finite, unramified) cover $\widetilde N$ of $N$, where $\pi:\
  \widetilde N \to N$ is the canonical surjection.


  So, first suppose that $\pi:\ \widetilde N \to N$ is a cover of $N$
  and that $\widetilde \Gamma$ is a simple cycle on $\widetilde N$. We
  must show that $\pi(\widetilde \Gamma)$ is a backtrackless, tailless
  cycle on $N$. Suppose otherwise, namely, that $(x, y, x)$ is part of
  the vertex sequence of $\pi(\widetilde \Gamma')$ for some
  $\widetilde \Gamma'$ equivalent to $\widetilde \Gamma$. Then it
  comes from $(\widetilde u, \widetilde v, \widetilde w)$ in
  $\widetilde \Gamma'$. In particular, this means that $v$ is adjacent
  to two distinct vertices $\widetilde u$ and $\widetilde w$ in
  $\widetilde N$, both of which project to $x$. This cannot happen in
  a finite unramified cover. Thus $\pi(\widetilde \Gamma)$ is
  backtrackless and tailless.

  For the converse, we must show that given a backtrackless, tailless
  cycle $\Gamma$ on $N$, there is a cover $\pi:\ \widetilde N \to N$
  and a simple cycle $\widetilde \Gamma$ on $\widetilde N$ lifting
  $\Gamma$. This is done by induction on the length of $\Gamma$, with
  cycles of length $3$, which are necessarily simple, providing the
  base case. For a nonsimple cycle $\Gamma$ of length greater than
  $3$, the idea is to break off the first simple cycle $\Gamma_1$
  appearing within $\Gamma$. Then $\Gamma$ is equivalent to a
  composition of $\Gamma_1$ with some other cycle $\Gamma_2$ which has
  length less than that of $\Gamma$. If $\Gamma_2$ is backtrackless,
  then it has a lift to a simple cycle by induction hypothesis and one
  must explicitly show how to ``glue together'' this lift with the
  cycle $\Gamma_1$ to form a simple lifting of $\Gamma$. The case
  where $\Gamma_2$ has backtracking presents a bit more difficulty,
  but is handled similarly.
\end{proof}


The following corollary is contained in the proof of
Th.~\ref{theorem:pseudos:in:cycle:1}.


\begin{corollary}\label{cor:to:pseudos:in:cycle}

  Consider the same setup as in Th.~\ref{theorem:pseudos:in:cycle:1}. The
  vector $p = (p_1,...,p_n) \in \N^n$ is an unscaled pseudo-codeword for $C$
  if and only if there is a backtrackless tailless cycle in $X$ which uses the
  $i^{\text th}$ edge exactly $p_i$ times for $1 \le i \le n$. Moreover, the
  unscaled pseudo-codewords of $C$ are in one-to-one correspondence with the
  monomials appearing with nonzero coefficient in the edge zeta function
  $\zeta_N$ of $N$. Finally, the Newton polyhedron of $\zeta_N$ (i.e.~the
  polyhedron spanned by the exponents of the terms in the Taylor series of
  $\zeta_N$) equals the fundamental cone $\fch{K}{H}$ of the code $C$.
\end{corollary}

\begin{itwreferences}

{\footnotesize


\bibitem{Tanner:81}
R.~M. Tanner, ``A recursive approach to low-complexity codes,'' {\em IEEE
  Trans.\ on Inform.\ Theory}, vol.~IT--27, pp.~533--547, Sept. 1981.

\bibitem{Koetter:Vontobel:03:1}
R.~Koetter and P.~O. Vontobel, ``Graph covers and iterative decoding of
  finite-length codes,'' in {\em Proc.\ 3rd Intern.~Conf.~on Turbo Codes and
  Related Topics}, (Brest, France), pp.~75--82, Sept.~1--5 2003.
\newblock Available online under \verb+http://www.ifp.uiuc.edu/~vontobel+.

\bibitem{Feldman:Karger:Wainwright:03:2}
J.~Feldman, D.~R. Karger, and M.~J. Wainwright, ``{LP} decoding,'' in {\em
  Proc.\ 41st Allerton Conf.~on Communications, Control, and Computing},
  (Allerton House, Monticello, Illinois, USA), October 1--3 2003.
\newblock Available online under
  \verb+http://www.columbia.edu/~jf2189/pubs.html+.

\bibitem{nicwi} N. Wiberg, ``Codes and Decoding on General Graphs'', Link\"oping Studies in Science and Technology, 
Ph.D thesis No. 440, Link\"oping, Sweden.
\verb+http://www.it.isy.liu.se/+ \verb+publikationer/LIU-TEK-THESIS-440.pdf+.

\bibitem{Kschischang:Frey:Loeliger:01}
F.~R. Kschischang, B.~J. Frey, and H.-A. Loeliger, ``Factor graphs and the
  sum-product algorithm,'' {\em IEEE Trans.\ on Inform.\ Theory}, vol.~IT--47,
  no.~2, pp.~498--519, 2001.

\bibitem{Hakimi:Bredeson:68:1}
S.~L. Hakimi and J.~Bredeson, ``Graph-theoretic error correcting codes,'' {\em
  IEEE Trans.\ on Inform.\ Theory}, vol.~IT--14, no.~4, pp.~584--591, 1968.

\bibitem{Forney:01:1}
G.~D. {Forney, Jr.}, ``Codes on graphs: normal realizations,'' {\em IEEE
  Trans.\ on Inform.\ Theory}, vol.~47, no.~2, pp.~520--548, 2001.

\bibitem{Massey:77:1}
W.~S. Massey, {\em Algebraic Topology: an Introduction}.
\newblock New York: Springer-Verlag, 1977.
\newblock Reprint of the 1967 edition, Graduate Texts in Mathematics, Vol. 56.

\bibitem{Stark:Terras:96:1}
H.~M. Stark and A.~A. Terras, ``Zeta functions of finite graphs and
  coverings,'' {\em Adv. Math.}, vol.~121, no.~1, pp.~124--165, 1996.

\bibitem{West:96:1}
D.~B. West, {\em Introduction to graph theory}.
\newblock Upper Saddle River, NJ: Prentice Hall Inc., 1996.

\bibitem{Hashimoto:89:1}
K.~Hashimoto, ``Zeta functions of finite graphs and representations of
  {$p$}-adic groups,'' in {\em Automorphic forms and geometry of arithmetic
  varieties}, vol.~15 of {\em Adv. Stud. Pure Math.}, pp.~211--280, Boston, MA:
  Academic Press, 1989.


}

\end{itwreferences}


\end{itwpaper}


\end{document}